\documentclass[twocolumn,aps,amssymb, prl]{revtex4}
\usepackage{epsfig}
\usepackage{graphicx}% Include figure files

\begin{document}

\title{Double-reversal thickness dependence of critical current in
superconductor-ferromagnet-superconductor Josephson junctions.}
\author{V. A. Oboznov, V.V. Bol'ginov, A. K. Feofanov, V. V. Ryazanov}
\email{ryazanov@issp.ac.ru}

\affiliation{Institute of Solid State Physics, Russian Academy of
Sciences, Chernogolovka, 142432, Russia}
\author{A. I. Buzdin}
\affiliation{{Institut Universitaire de France, Paris and CPMOH, UMR
5798, University Bordeaux I, 33405 Talence, France}}
\date{\emph{\today}}

\begin{abstract}
We report the first experimental observation of the two-node
thickness dependence of the critical current in Josephson junctions
with a ferromagnetic interlayer. Vanishings of the critical current
correspond to transitions into $\pi$-state and back into
conventional $0$-state. The experimental data allow to extract the
superconducting order parameter oscillation period and the pair
decay length in the ferromagnet. We develope a theoretical approach
based on Usadel equations, which takes into account the spin-flip
scattering. Results of numerical calculations are in good agreement
with the experimental data.
\end{abstract}

\maketitle

%\draft

One of the exciting topics in studying the coexistence of
superconductivity ($S$) and ferromagnetism ($F$) is
proximity-induced sign-reversal superconductivity in a ferromagnet
close to an $SF$-interface \cite{Buz82,Buz92}. The superconducting
order parameter does not simply decay into the ferromagnet but also
oscillates. An undoubted evidence of sign-reversal spatial
oscillations of the superconducting order parameter in a ferromagnet
was the observation of the $\pi$-state in $SFS$ Josephson junctions
\cite{PRL,Aprili,Sellier,JLTP}. '$\pi$-junctions' \cite{Bul77} are
weakly coupled superconducting structures which demonstrate
$\pi$-shift of the macroscopic phase difference in the ground state.
The relation between the superconducting current $I_s$ and the phase
difference $\varphi$ in a Josephson junction is described by a
$2\pi$ -periodic function. In the simplest case of a tunnel barrier
or a barrier, made of dirty normal metal, one finds $I_s=I_c\sin
\varphi$. The Josephson $\pi$-junction has an anomalous
current-phase relation $I_s=I_c\sin (\varphi + \pi) =-I_c\sin
\varphi $, i.e. it is characterized (nominally) by the negative
critical current~\cite{Bul77}. Spatial oscillations of the
superconducting order parameter in a ferromagnet close to an
$SF$-interface was predicted in Ref.~\cite{Buz92}. A physical origin
of the oscillations is the exchange splitting of the spin-up and
spin-down electron subbands in a ferromagnet. It was discussed in
Refs.~\cite{PRL,Aprili,Sellier,JLTP} that in order to observe
manifestations of the transition into the $\pi$-state one should
fabricate an $SFS$ sandwich with the $F$-layer thickness $d_F$ close
to integer numbers of half-periods of the order parameter spatial
oscillations $\lambda_{ex}$/2. The period is $\lambda_{ex}=2\pi
\xi_{F2}$, where the oscillation (or "imaginary") length $\xi_{F2}$
can be extracted from the complex coherence length $\xi_F$ in a
ferromagnet: $\frac 1 {\xi_{F}}=\frac 1{\xi_{F1}}+i\frac 1
{\xi_{F2}}$. In the simplest case the imaginary length $\xi_{F2}$
and the order parameter decay length $\xi_{F1}$ are equal
\cite{Buz92}: $\xi_{F1}=\xi_{F2}=\sqrt{\hbar D/E_{ex}}$, where $D$
is the diffusion coefficient for electrons in a ferromagnet and
$E_{ex}$ is the exchange energy responsible for sign-reversal
superconductivity in a ferromagnet. Temperature changes of the
coherence length related to the thermal energy contribution to
pair-breaking processes were introduced in Ref. \cite{PRL}, in which
temperature $0-\pi$-transition was observed for the first time. The
expressions for $\xi_{F1}$ and $\xi_{F2}$ are the following:
\begin{equation}
\xi_{F1,2} = \sqrt{\frac{\hbar D}{\sqrt{(\pi k_BT)^2 + E_{ex}^2}\pm
\pi k_BT}}\simeq \sqrt{\frac{\hbar D}{E_{ex}}}(1\mp \frac{\pi
k_BT}{2E_{ex}}), \label{xiT}
\end{equation}
The latter approximation corresponds to the case $E_{ex}\gg k_BT$,
which is valid for experiments discussed below.

Detailed experimental studies of the critical current thickness
dependence for $Nb-Cu_{0.47}Ni_{0.53}-Nb$ Josephson junctions has
been started by us in Ref.~\cite{JLTP}. A very large decay of the
critical current and its sharp reentrant behavior for thicknesses
close to 23 nm have been observed. An analysis of the experimental
data and their comparison with the modern model described below have
shown that the observed deep minimum is probably the reverse
transition from the $\pi$- into the $0$-state at the $F$-layer
thickness close to full oscillation period while the first node of
the dependence has to be at the thickness of about 10 nm. Thus, the
presented work is devoted to finding of the two-node behavior of the
$SFS$ junction critical current as well as to discussion of
mechanisms of the strong order parameter decay in a ferromagnetic
$CuNi$ interlayer.

\begin{figure}[tbp]
\begin{center}
\includegraphics[width=0.4\textwidth, clip]{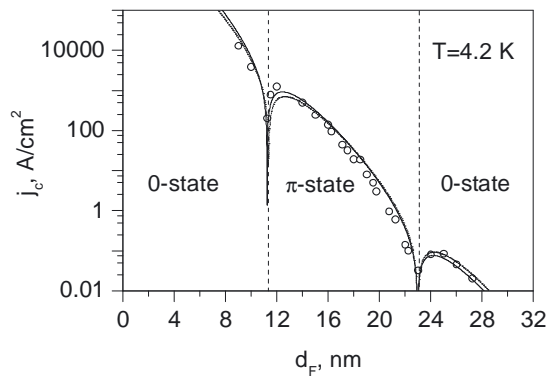}
\end{center}
\caption{The F-layer thickness dependence of the critical current
density for $ Nb-Cu_{0.47}Ni_{0.53}-Nb$ junctions at temperature
$4.2~K$. Open circles present experimental results, solid and dashed
lines show model calculations discussed in the second part of the
paper.} \label{IdF}
\end{figure}

\begin{figure}[tbp]
\begin{center}
\includegraphics[width=0.4\textwidth, clip]{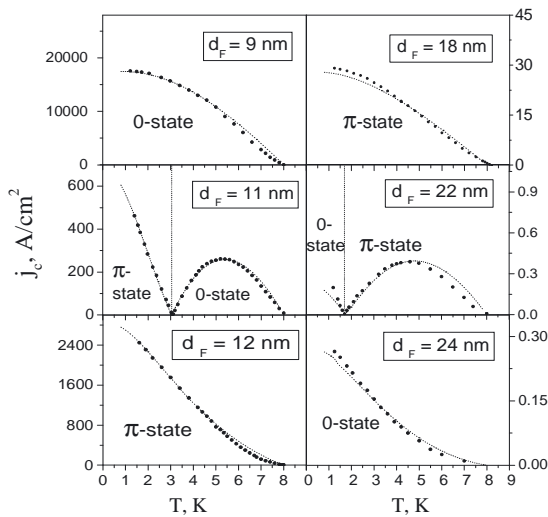}
\end{center}
\caption{Temperature dependences of the $SFS$ junction critical
current density at several $F$-layer thicknesses close to the
critical ones. The dashed lines show calculation results based on
Eq. (\protect\ref{Ic}).} \label{IT}
\end{figure}

In fact a nonmonotonic $I_{c}(d_{F})$ dependence close to
$0-\pi$-transition was observed for the first time in
Ref.~\cite{PRL} and has been presented there as a number of
$I_{c}(T)$ curves for different thicknesses $d_{F}$. Later Kontos et
al~\cite{Aprili} for $Nb-Pd_{0.9}Ni_{0.1}-Nb$ and then Sellier et
al~ \cite{Sellier} for $Nb-Cu_{0.52}Ni_{0.48}-Nb$ junctions measured
detailed reentrant $I_{c}(d_{F})$ curves for $F$-interlayer
thicknesses close to $ 0-\pi $-transition. In this work we have
investigated the thickness dependence of the $SFS$ junction critical
current density in a wide thickness range for sandwiches fabricated
as described in Ref.~\cite {JLTP}. All junctions had their lateral
sizes smaller than the Josephson length and uniform current
distribution. To do this the junctions with F-layer thicknesses of
less than 17 nm were made with the contact area $10\times 10~\mu
m^{2}$ and all the rest had the area $50\times 50~\mu m^{2}$.
Weakly-ferromagnetic $Cu_{0.47}Ni_{0.53}$-interlayers had the Curie
temperature of about 60 K. In the thickness interval $8-28\,nm$ we
had about 6 orders of the critical current density change with
vanishings at two $d_{F}$ values as it is presented in
Fig.~\ref{IdF}.
One can see that undoubtly the curve  demonstrates both direct
$0-\pi $-transition and reverse transition from $\pi$- to $0$-state.
In transition points the critical current $I_{c}(d_{F})$ is equal to
zero and then should formally change its sign. Since in real
experiments we could measure only magnitude of the critical current,
the dependence $I_{c}(d_{F})$ between two sharp cusps is the
negative (corresponding to the $\pi $-state) branch of the curve
which is reflected into the positive region. Due to slight
temperature dependence of the order parameter oscillation period in
our weak ferromagnet (described by (\ref{xiT})) we could pass
through the transition points using samples with critical $F$-layer
thicknesses 11 nm and 22 nm by means of temperature decrease.
Temperature $0-\pi$- and $\pi-0$ -transitions are presented in
middle panels of Fig.~\ref{IT}. Upper and lower panels show the
critical current temperature behavior for samples with close F-layer
thicknesses. One can see that we lost a possibility to detect
temperature transitions changing the thickness only by $1-2 nm$.
This implies that the temperature decrease from $9~K$ down to $1~K$
is accompanied by the decrease of $1-2~nm$ in the spatial
oscillation period and by the decrease of about $0.3~nm$ in the
oscillation length. In this temperature range the change of
$\xi_{F1}$ is about $0.2~nm$ as it has been estimated from
$I_c(d_F)$ curves at different temperatures. At the same time simple
evaluations of $\xi _{F1}$ (obtained from the slope of the
$I_{c}(d_{F})$ envelope) and $\xi _{F2}$ (estimated from the
interval between two minima) show a large difference between these
two lengths (1.3 nm and 3.5 nm, correspondingly) that can not be
explained by the thermal contribution described by (\ref{xiT}).

So, to carry out a theoretical analysis of the results obtained, we
need to specify the nature of additional depairing processes that
increase $\xi _{F2} $ and decrease $\xi _{F1}$. As the $F$ layer is
an alloy, a role of the magnetic scattering may be quite
important~\cite{Sellier,JLTP}. Magnetic inhomogeneity is related
above all to $Ni$-rich clusters \cite{Mills,Houghton} arising in
$Cu_{1-x}Ni_{x}$ ferromagnet for $x$ close to 0.5. In the region of
these concentrations when the Curie temperature is small, we may
expect that the inverse spin-flip scattering time $\hbar \tau
_{s}^{-1}$ could be of the order of the average exchange field
$E_{ex}$ or even larger. This circumstance strongly modifies the
proximity effect in the $SF$ systems. A role of spin-orbit
scattering should be neglected for the $CuNi$ alloy since it is
substantial only in ferromagnets with large atomic number $Z$. To
take into account the exchange field and the magnetic scattering in
the framework of Usadel equations it is necessary just to substitute
Matsubara frequencies by $\omega \rightarrow \omega +iE_{ex}
+G\hbar/\tau _{s}$ \cite{Buzdin85}, where $G$ is the normal Green's
function. Note that this procedure assumes a presence of the
relatively strong uniaxial magnetic anisotropy which prevents mixing
of spin-up and spin-down Green functions \cite{Buz2003}.

To have some idea about the influence of the magnetic scattering on
the proximity effect we may start with the linearized Usadel
equation \cite{usad70} for the anomalous Green's function in a
ferromagnet

\begin{equation}
\left( \left| \omega \right| +iE_{ex}sgn\left( \omega \right)
+\frac{\hbar}{\tau _{s}}\right) F_{f}-\frac{\hbar
D}{2}\frac{\partial ^{2}F_{f}}{\partial x^{2}}=0.  \label{Us + magn
scatt}
\end{equation}

The exponentially decaying solution has the form
\begin{equation}
F_{f}\left( x,\omega >0\right) =A\exp \left(
-x(k_{1}+ik_{2})\right),
\end{equation}

with

$$k_{1}=\frac{1}{\xi _{F}}\sqrt{\sqrt{1+\left( \frac{\omega}{E_{ex}} +
\frac{\hbar}{E_{ex}\tau _{s}}\right) ^{2}}+\left(\frac{
\omega}{E_{ex}} +\frac{\hbar}{E_{ex}\tau _{s}}\right) },
$$

$$
k_{2}=\frac{1}{\xi_{F}}\sqrt{\sqrt{1+\left( \frac{\omega}{E_{ex}} +
\frac{\hbar}{E_{ex}\tau _{s}}\right) ^{2}}-\left(\frac{
\omega}{E_{ex}} +\frac{\hbar}{E_{ex}\tau _{s}}\right)}.
$$

Here $\xi _{F}=\sqrt{\hbar D/ E_{ex}}$ and $\xi_{F1,2}=1/k_{1,2}$.
The anomalous Green's function $F_{f}$ at $\omega \sim k_B T_{c}$
gives us an idea about the spatial variation of the Cooper pair wave
function. In the limit of the vanishing magnetic scattering and $k_B
T_{c}<< E_{ex}$ the decaying $(\xi _{F1})$ and the oscillating $(\xi
_{F2})$ lengths are practically the same. However, if the spin-flip
scattering time becomes relatively small $ E_{ex}\tau _{s}/\hbar
\lesssim 1$, the decaying length could be substantially smaller than
the oscillating length. This results in much stronger decrease of
the critical current in $SFS$ junctions with increase of the $F$
layer thickness.

We have seen it experimentally~\cite{JLTP} that the form of
$I_{c}(d_{F})$ dependence varies a little with temperature, so a
good idea about this dependence may be already obtained from the
temperature region near $T_{c}$. Using $k=k_1+ik_2$ in the form
\begin{equation}
k=\sqrt{2\left( \left\vert \omega \right\vert +iE_{ex}sgn\left(
\omega \right) + \frac{\hbar}{\tau _{s}}\right)/\hbar D}
\end{equation}
we can obtain (see Ref.~\cite{Buz2003}) for the case of good
$SF$-interface transparency and $d_{F}\gg \xi _{F1}$ the following
expression for the critical current:
\begin{equation}
j_{c}\sim e^{-d_{F}/\xi _{F1}}\left( \cos {d_{F}/\xi _{F2}}+(\xi
_{F1}/\xi _{F2})\sin {d_{F}/\xi _{F2}}\right), \label{Ic_SFS_Buzdin}
\end{equation}
where $\xi_{F1,2}$ are taken in the limit of $\omega \ll E_{ex},
\hbar/\tau_s$.

Now we shall address the question of the exact thickness and
temperature dependence of the critical current in $SFS$ junctions.
To deal with the complete set of the Usadel equations it is
convenient to apply the usual parametrization of Green's functions :
$G_{f}=\cos \Theta (x)$ and $F_{f}=\sin \Theta (x)$. Then for
$\omega >0$ the Usadel equation is written as

\begin{equation}
\left(\omega +iE_{ex}+\frac{\hbar\cos \Theta }{\tau _{s}}\right)
\sin \Theta -\frac{\hbar D}{2}\frac{\partial ^{2}\Theta }{\partial
x^{2}}=0. \label{Us nonlin}
\end{equation}

If the temperature variation of the exchange field is negligible at
$T<T_{c}$ the most direct way how the temperature could interfere is
through the Matsubara frequencies. The presence of the magnetic
scattering provides another mechanism of the critical current
temperature dependence - through the normal Green's function
$G_{f}=\cos \Theta.$  The important range of the Matsubara
frequencies variation is of the order of $k_B T_{c}$ for
superconductivity. Then in the case of the relatively strong
magnetic scattering $\hbar\tau _{s}^{-1}>>k_BT_{c}$ the second
mechanism of the temperature dependence will be predominant.

In the limit of relatively large $F$-layer thicknesses $d_{F}>\xi
_{F1}$ and rigid boundary conditions we may obtain an analytical
solution of Eq.(\ref{Us nonlin}) and the expression for the critical
current density reads

\begin{equation}
j_{c}(d_F,T)=\frac{64\sigma _{n}\pi k_BT_{c}}{e\xi _{F}}
%TCIMACRO{\func{Re}}%
%BeginExpansion
\mathop{\rm Re}%
%EndExpansion
\left( \stackrel{\infty }{%
%TCIMACRO{\underset{n>0}{\sum }}%
%BeginExpansion
\mathrel{\mathop{\sum }\limits_{n>0}}%
%EndExpansion
}\frac{{\cal F}\left( n\right) q\exp (-qy)}{\left[ \sqrt{\left(
1-p^2\right) {\cal F}\left( n\right) +1}+1\right] ^2}\right)
\label{Ic}
\end{equation}
with the function
$$
\mathcal{F}\left( n\right) =\frac{\left( \Delta /\left( 2\pi
k_BT\right) \right) ^2}{\left[ n+1/2+\sqrt{\left( n+1/2\right)
^2+\left( \Delta /\left( 2\pi k_BT\right) \right) ^2}\right] ^2},
$$
and $y=d_F/\xi_F$, $q=\sqrt{2i+2\alpha +2 \widetilde{\omega }}$ ,
where $\alpha =\hbar /(\tau _{s}E_{ex})$, $\widetilde{\omega
}=\omega /E_{ex}=\frac{2\pi (n+1/2)(T/T_c)}{E_{ex}/k_B T_c}$ and
$1-p^2=(i+\widetilde{\omega })/(\alpha +i+\widetilde{\omega }). $

In the limit $\alpha \rightarrow 0$ and $k_BT_{c}<<E_{ex}$ Eq.
(\ref{Ic}) coincides with that obtained previously in Ref.
\cite{Buzdin91}. The theoretical fit of our experimental results
which is based on Eq. (\ref{Ic}) is presented in Fig.~\ref{IdF} by
the solid line and in Fig.~\ref{IT} by dashed lines. Besides the
dashed line in Fig.~\ref{IdF} shows calculations made using Eq.
(\ref{Ic_SFS_Buzdin}). One can see a good agreement obtained with
the following parameters: $E_{ex}/k_B\approx 850~K$,
$\hbar/\tau_s\approx 1.33~E_{ex}$, $\xi_F=2.16~nm$. The fitting also
yields considerable value of 'dead' layers $d_0$:
$2d_0\approx4.3~nm$, which do not take part in creating of the
'sign-reversal' superconductivity. The dead layer may arise due to
not full correspondence of the theoretical approach and the real
system. On the other hand, other experiments \cite{Courtois} also
demonstrate the existence of large enough (2-3 nm) nonmagnetic
layers at $SF$-interfaces.

\begin{figure}[tbp]
\begin{center}
\includegraphics[width=0.4\textwidth, clip]{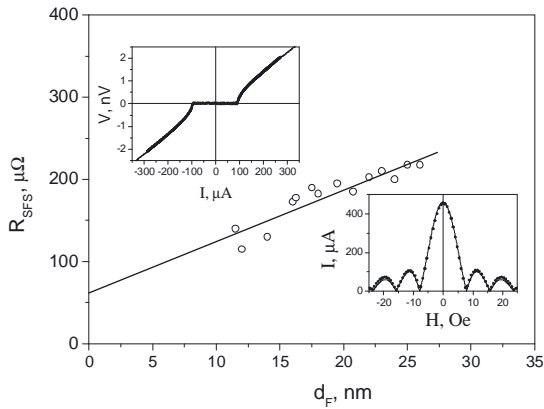}
\end{center}
\caption{Resistance of $SFS$-sandwiches normalized to the junction
area of $ 10\times10~\mu m^2$ vs. the $F$-layer thickness. Insets
show typical IV- and Fraunhofer ($I_c(H)$) dependences of $SFS$
junctions with ideal fitting by well-known Josephson expressions.}
\label{Curves}
\end{figure}

A final remark concerns the real transparency of $SF$-interfaces in
our $SFS$ sandwiches. In modern theories an interface transparency
is characterized by a parameter $\gamma_B = (R_B S/\rho_F \xi^*)$,
where $R_B$ is interface resistance per unit area, $S$ is the $SFS$
junction area, $\rho_F$ is $F$-layer resistivity and
$\xi^*=\sqrt{\hbar D/2\pi k_BT_c}$. To estimate $R_B$ and $\rho_F$
we have carried out detailed measurements of $SFS$-junctions
$IV$-characteristics. The upper inset in Fig.~ \ref{Curves} shows
that $IV$-characteristics are described by the expression
$V=R\sqrt{I^2-I_c^2}$.
The linear approximation presented in Fig.~\ref{Curves} has given
$R_B\approx 30~\mu\Omega$ for junctions with the area of
$10\times10~\mu m^2$ and $\rho_F\approx 62~\mu\Omega\cdot cm$. It
allows to estimate following ferromagnet parameters: the electron
mean free path $l\approx$ 1 nm, the diffusion coefficient
$D\approx5.2~cm^2/c$ and the characteristic spatial scale
$\xi^*\approx9.4~nm$. Values obtained determine the good enough
transparency parameter $\gamma_B=0.52$ that confirms the validity of
the approximation used.

An additional breakthrough of the work is fabrication of
$\pi$-junctions with large enough critical current density. Solving
of this problem has enabled detailed experimental investigations of
$0-\pi$-transition peculiarities, reliable detections of second
harmonic in the current-phase relation \cite{Sellier2004} and the
$0-\pi$-coexistence \cite{Radovich}. High magnitude of the critical
current also allows to use $SFS$ $\pi$-junctions as stationary phase
$\pi$-shifters in novel modifications of the digital and quantum
logic \cite{appl}. In proposed logic circuits $\pi$-junctions are
connected together with ordinary tunnel junctions and should not
introduce themselves any noticeable phase shift during dynamical
switchings in the rest of the circuit. This is possible only if the
$\pi$-junction critical current is much larger then critical
currents of other junctions. The $Nb-CuNi-Nb~\pi$-junctions are
based on the standard niobium thin film technology so they can be
incorporated directly into existing architectures of the
superconducting electronics.

Thus, both $0-\pi$ and reverse $\pi-0$ transitions have been
detected in $SFS~(Nb-CuNi-Nb)$ junctions for the first time. The
double-reversal thickness dependence of the critical current is a
most striking evidence  of the superconducting order parameter
spatial oscillations in a ferromagnet close to $SF$-interface. We
have also observed that the oscillation length in the ferromagnetic
$CuNi$ alloy is considerably larger than the the pair decay length.
We have presented a theoretical description of an extra mechanism of
the order parameter decay in $CuNi$, mainly related to the strong
spin-flip scattering on magnetic inhomogeneity.

We are grateful to A.Bobkov, I. Bobkova, Y. Fominov, A. Golubov, M.
Kupriyanov and A. Rusanov for helpful discussions and to N. S.
Stepakov for assistance during experiments. This work was supported
by Russian Foundation for Basic Research, Programs of Russian
Academy of Sciences, INTAS(grant no. 01-0809) and partially by ESF
"Pi-shift" Programme and French ECO-NET 2005 Programme".

\end{document}